\documentclass{article}
\usepackage{graphicx} % Required for inserting images
\usepackage{amsmath}
\usepackage{color}

\title{Tracking the dynamical, chemical and spectral evolution of molecular cores with PrestaLine: Gorynych}
\author{Olga Kochina, Dmitri Wiebe,\\
Yaroslav Pavlyuchenkov, Maria S. Kirsanova}
\begin{document}

\maketitle
\begin{abstract} 
\textbf{~}We present \textit{PrestaLine: Gorynych}, a comprehensive numerical tool designed to model the dynamical, chemical, and spectral evolution of collapsing molecular cores from the prestellar phase to protostellar accretion. The code integrates three key components: (1) \textit{Kamelung}, a 1D hydrodynamics module simulating gravitational collapse up to first hydrostatic core (FHSC) formation followed by an accretion of envelope onto a young star; (2) \textit{Presta}, a chemical evolution module post-processing density and temperature profiles to compute time-dependent molecular abundances; and (3) \textit{Uran(IA)}, a radiative transfer module generating synthetic molecular line spectra. We apply \textit{Gorynych} to compare low-mass (5~M\(_\odot\)) and high-mass (50~M\(_\odot\)) cores, finding that their dynamical evolution is remarkably similar, with differences primarily in spatial scaling. Chemical evolution reveals sharp abundance changes during the FHSC transition, particularly for CO, H\(_2\)O, and HCO\(^+\), though pre-collapse chemical initialization has minimal impact on most species. Spectral maps of \(^{13}\)CO(2--1), HCO\(^+\)(3--2), and H\(_2\)O(1\(_{10}\)--1\(_{01}\)) lines show distinct kinematic signatures of infall and depletion, with high-mass cores exhibiting spatially extended emission. Our results highlight \textit{Gorynych}'s utility to couple theoretical collapse models and observations, providing a framework to diagnose core evolution and initial conditions from molecular line data.
\end{abstract}

\section{Introduction}

The very beginning of the way for interstellar matter to become stellar, forming a protostar in a molecular cloud, is a complicated phenomenon that engages many different processes and scales. Observations of molecular lines are a primary source of information about these processes. To properly interpret these observations in terms of an evolutionary status and physical properties of a star-forming object, we need to invoke models of three kinds.
Modelling of the chemical evolution is required to compute chemical abundances. Modelling of radiative transfer allows converting these abundances into corresponding line intensities, to relate them to observations directly. Finally, modelling of the object dynamics provides information on both the evolution of its physical structure, required by the chemical part, and velocities, needed for correct calculation of spectra.

This problem can be solved in a number of ways. The most comprehensive way is to consider chemistry and dynamics simultaneously, obtaining self-consistent spatial distributions of densities, velocities, and chemical abundances \cite{1997MNRAS.292..601S}, which then can be used as an input for a radiation transfer model \cite{2003ARep...47..176P,2019MNRAS.486.2525K}. However, this approach is very computationally expensive, which implies either simplified chemistry \cite{2016A&A...587A..60F}, or simplified dynamics \cite{2018A&A...615A..15S}, or both. An alternative way is to use a so-called post-processing, when a dynamical model, which may be quite complicated, is computed first, and then equations of chemical kinetics are solved on the top of the obtained distributions of density, temperature, and other parameters. This method lacks self-consistency, which may be warranted to simulate heating and cooling rates, depending on abundances of certain molecules (CO, water, etc.), or a coupling between matter and magnetic field, depending on molecular ion abundances.

These approaches have been implemented in a number of studies with a special emphasis on a transition from a prestellar phase to a protostar and to a protoplanetary disk. Furuya et al. \cite{2012ApJ...758...86F} have used a 3D radiation hydrodynamics simulation to follow a chemical evolution of a parcel moving along the flow until a first hydrostatic core (FHSC) stage is reached. Overall, $10^4$ trajectories have been considered. A similar approach has also been utilized in \cite{2013ApJ...775...44H} and \cite{2020A&A...643A.108C}. A more detailed post-processing study of a specific collapsing core extracted from a large-scale 3D simulation has been performed by Hincelin et al. \cite{2016ApJ...822...12H}, who have followed the evolution of the collapsing protostar till the formation of a first hydrostatic core.

Jin et al. \cite{2022ApJ...935..133J} have adopted a 1D model of the FHSC formation. This model was post-processed with a so-called three-phase chemical model, which considers two phases of dust icy mantles (surface and bulk) along with the gas phase. A special emphasis of this work was on dust mantle processes, including non-diffusive chemistry \cite{2020ApJS..249...26J}, in order to make some predictions for the JWST observations.

Here we present a tool that provides all the modelling ingredients required to create a consistent picture of the early stages of star-formation. PrestaLine: Gorynych is a package that simulates the collapse, allowing us to track the changes in physical structure of the object up to the formation of a protostar and further on. The chemical evolution is modelled taking into account the evolution of the physical structure and a wide range of chemical processes and reactions. The comprehensive package for calculation of the radiative transfer makes predictions easy to refer to and explain existing observations. 

In this work we implement Gorynych to provide the comparison study of the evolution of the molecular clouds of two different masses, examine the impact the initial mass has on collapse and chemical composition of the object, compare synthetic spectra for different evolutionary stages, and evaluate the need to take into account an ``evolved'' initial chemical composition.

The description of the PrestaLine: Gorynych tool and the physical models used for the calculations is presented in Section 2. Results of calculations are described in Section 3 and discussed in Section 4.

\section{Physical model, methods and numerical code overview}

The PrestaLine project started back in 2019 as a tool to model chemical evolution and molecular line parameters for diffuse, translucent, and dense interstellar medium \cite{2020INASR...5..294K}. It is aimed to provide a direct comparison of theoretical models with observations in order to constrain parameters of prestellar and protostellar objects and their chemical ages. Its first version comprised two large separate codes, Presta for modeling chemistry (developed
by our group \cite{2013ARep...57..818K}) and publicly available numerical package RADEX for radiative transfer \cite{2007A&A...468..627V}, accompanied by visualization tools.

Presta plus RADEX complex has proved its efficiency reproducing and explaining the DR21(OH) molecular subcores observations \cite{2022AstL...48..517L}, but further studies warranted a more detailed approach. PrestaLine in its initial version adopted quite a simplified description of physical parameters within an object, which has been divided into three static subregions. Physical evolution of the object has not been taken into account.

Now we present an updated version of PrestaLine, which includes an evolutionary model of a spherically symmetric protostellar molecular core. Calculations of molecular spectra are performed with a two-dimensional radiative transfer code, taking into account the detailed physical structure of the modelled object. The new version is named Gorynych, after the three-headed dragon from the Eastern Slavic folklore, as it similarly unites three equally important components. These are Kamelung, a code that calculates collapse of a molecular core, Presta for chemical evolution, and the radiative transfer code Uran(IA) with the visualization and analysis tools. The codes work in a post-processing manner and have no on-the-fly communication with each other, being linked by output files, used as an input for the next stage.

A first step of the processing is done by Kamelung (see subsection 2.1 for details), a 1D code, which simulates the collapse of the cloud up to the formation of the FHSC and the subsequent accretion phase. It utilizes hydrodynamics and radiative transfer to calculate the physical and thermal evolution of the cloud. It separately models the processes of heating and cooling of gas and dust considering the radiation of the protostar and the external radiation field. The cloud is divided into a set of concentric cells, moving along the radius as the object evolves. At a certain moment, a central part is removed from the computational domain, whic is an onset of the accretion phase. During that phase, the cells fall onto the protostar, disappearing from the calculation, gradually depopulating the set.

The saved history of the evolving cloud is then converted into input files used by Presta, a model of chemical evolution also adopting the spherical symmetry. Presta (see subsection 2.2) is able to treat complex dust chemistry, taking into account multiple dust populations and chemistry in the grain mantles. Presta calculates chemical evolution on top of the pre-computed physical evolution, processing the motion of the cells during both the prestellar and accretion phases. The result of the Presta modeling can be presented in the form of column densities and radial abundance profiles.  

Information on the abundance distribution and physical parameters within the cloud is used further as an input for the third step, radiative transfer, provided by Uran(IA) (see subsection 2.3). Uran(IA) allows one to build emission spectra in a form consistent with the characteristics of the observed ones. Theoretical spectral maps can be convolved with a typical telescopes beam pattern. Uran(IA) adopts the results on self-consistent level populations obtained by URAN, a two-dimensional numerical code that implements Monte Carlo methods. URAN uses a spherical coordinate system, so it works effectively with our one-dimensional problems. The results of Uran(IA) work can be presented in various forms, such as spectral maps, integrated intensity maps, channel velocity maps, position-velocity diagrams, etc.

Gorynych works in a post-processing mode, keeping each ``dragon head'' independent in its functionality, allowing individual updates and upgrades of parts without hampering the stability of the system as a whole. Next three subsections provide deeper details on each of the heads of the Gorynych package.

\subsection{Head 1: Kamelung}

The numerical code Kamelung was described in detail in \cite{2013ARep...57..641P,2015ARep...59..133P}. Here we provide only its brief overview and describe the modifications relevant for the current study. The physical model is based on a numerical solution of the hydrodynamical and radiative transfer equations:
\begin{eqnarray}
&&\dfrac{\partial \rho}{\partial t} + \nabla\left(\rho
{\mathbf{v}}\right) = 0,
\label{hd01}\\
&&\dfrac{\partial {\mathbf{v}}}{\partial t} + \left({\mathbf{v}}
\cdot \nabla\right) {\mathbf{v}} = {-}\dfrac{\nabla p}{\rho} + {\mathbf{g}},
\label{hd02}\\
&&\dfrac{\partial E_{\textrm{g}}}{\partial t} + \nabla \left(
E_{\textrm{g}} {\mathbf{v}} \right) = - p  \nabla {\mathbf{v}}
+ \Gamma_{\textrm{g}} - \Lambda_{\textrm{g}} -
\Lambda_\textrm{gd},
\label{hd03} \\
&&\dfrac{\partial E_{\textrm{d}}}{\partial t}+ \nabla \left(
E_{\textrm{d}} {\mathbf{v}} \right)=
-\sigma_P(aT_{\textrm{d}}^4 - E_{\textrm{r}}) +
\Lambda_\textrm{gd} + S,
\label{hd04} \\
&&\dfrac{\partial E_{\textrm{r}}}{\partial t} - \nabla
\left(D_\textrm{F}\nabla E_{\textrm{r}}\right) =
\sigma_P(aT_{\textrm{d}}^4 - E_{\textrm{r}}). 
\label{hd05}
\end{eqnarray}
Here, $\rho$ is the total (gas+dust) density, ${\mathbf{v}}$ is the velocity, $p$ is the pressure, $\mathbf{g}$ is the gravitational acceleration, $E_{\textrm{g}}$ and $E_{\textrm{d}}$ are the thermal energy (per unit volume) of gas and dust, $E_{\textrm{r}}$ is the energy of thermal radiation, $\Gamma_{\textrm{g}}$ is the gas heating function, $\Lambda_{\textrm{g}}$ is the gas cooling function, $\Lambda_\textrm{gd}$ is the rate of thermal energy exchange between gas and dust, $S$ is the rate of dust heating by interstellar radiation, $T_{\textrm{d}}$ is the dust temperature, $a$ is the radiation constant, $D_\textrm{F}$ is the radiation diffusion coefficient in the FLD approximation, $\sigma_P=c\rho_\textrm{d}\kappa_\textrm{P}$, $c$ is the speed of light, $\rho_\textrm{d}$ is the dust density$, \kappa_\textrm{P}$ is the Planck dust opacity.

Equations~\eqref{hd01} and \eqref{hd02} are continuity and Euler equations, Eq.~\eqref{hd03} and \eqref{hd04} describe the advection and evolution of gas and dust thermal energy, Eq.~\eqref{hd05} provides the evolution of thermal radiation energy. Altogether, these equations are specified to follow the evolution of a self-gravitating molecular core with a detailed balance between heating and cooling. The dust and gas temperatures are calculated separately taking into account exchange of thermal energy via collisions. During the collapse, the molecular core is heated by the pressure force as well as by cosmic rays and interstellar radiation. The cooling is due to escape of molecular and dust thermal radiation.

Eqs.~(\ref{hd01}--\ref{hd05}) are solved in a spherically symmetric configuration using Lagrangian coordinates. To solve the system we use method of splitting into physical processes (hydrodynamics and radiative transfer subsystems) with implicit finite difference schemes for both problems.

\subsubsection{Non-uniform Lagrangian grid}

In the original version of the Kamelung code \cite{2015ARep...59..133P}, all Lagrangian cells were equal in mass. However, this does not provide a good resolution either near the border of the FHSC at the prestellar phase or near the boundary of the sink cell at the accretion phase. Thus, we introduced the non-uniform Lagrangian grid which allows better resolving these regions.  The cumulative mass $M(k)$ of the first $k$ cells, i.e. $M(k) = \sum\limits_{i=1}^{k} m(i)$, where $m(i)$ is the mass of $i$-th cell), is set to
\begin{equation}
    M(k) = M_\text{core}\left(\dfrac{k}{N}\right)^{\beta},
\end{equation}
where $N$ is the total number of cells while the exponent $\beta$ is typically chosen to be 1.5.

\subsubsection{Molecular cooling}

A cooling function $\Lambda(T)$ in a collapsing interstellar core is a complicated function of its chemical composition, which in its own turn depends on the local temperature \textbf{(Eq.\ref{cool1})}. Thus, in an ideal model, one would need to compute the cooling rate self-consistently with the chemical composition and dynamical evolution (see e.g. \cite{1997MNRAS.292..601S,2009ARep...53..611K}). But this would make our computation intractable, so we have employed a somewhat different approach, similar to the one described by Goldsmith \cite{2001ApJ...557..736G}. First, we have estimated the cooling function for a range of expected physical conditions using publicly available codes Meudon PDR\footnote{http://ism.obspm.fr} \cite{2006ApJS..164..506L} and version 23.01 of Cloudy, last described by \cite{2023RMxAA..59..327C,2023RNAAS...7..246G}. Then, we adopted a bi-linear approximation
\begin{equation}
    \lg \Lambda_\text{mol} = a \lg T + b \lg n +c, \label{cool1}
\end{equation}
where $a=2.46$, $b=1.22$, $c=-29.9$. This formula fits the pre-computed cooling functions and is close to Goldsmith's representation in the considered density range $n_\text{min}<n<n_\text{max}$, where $n_\text{min}=10^2$~cm$^{-3}$ and $n_\text{max}=10^7$~cm$^{-3}$. When hydrogen density is below $n_\text{min}$ and above $n_\text{max}$ we calculate $\Lambda_\text{mol}$ using Eq.~\eqref{cool1} with $n_\text{min}$ and $n_\text{max}$, correspondingly.

\begin{figure}
\includegraphics[width=\textwidth]{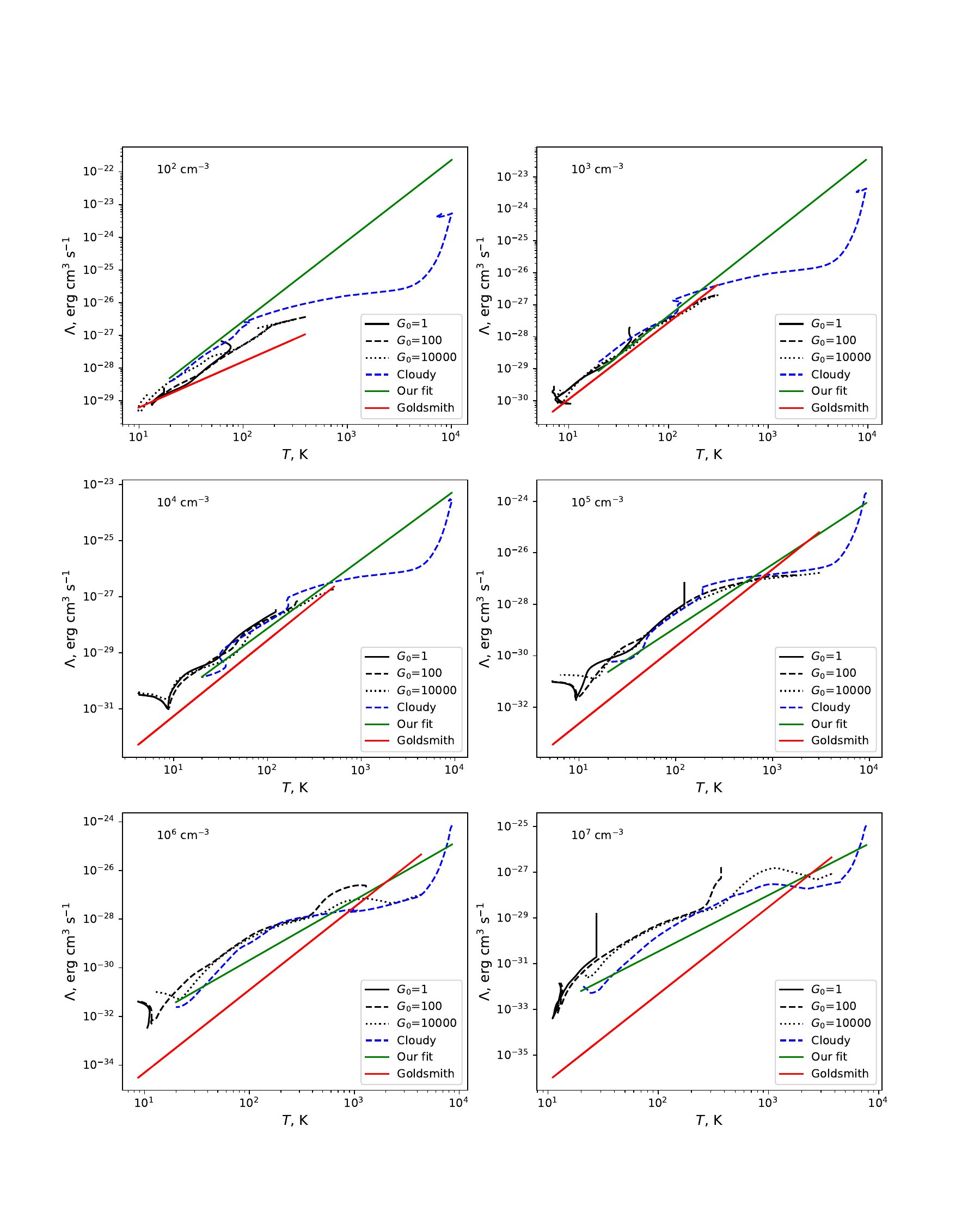}
\caption{Various cooling functions and the adopted approximation.}
\label{coolfun}
\end{figure}

In Fig.~\ref{coolfun} we present cooling functions computed with Meudon PDR and Cloudy 23.01. Meudon curves (black) are computed for various values of illuminating UV field intensity ($G_0$) as indicated in the legend. Cloudy curves (blue) are computed under assumption of an O~star ($T_{\rm eff}=50000$\,K) illuminating a molecular cloud with a uniform density distribution from 3 pc distance. This setup does not correspond to the physical situation considered in the paper and only serves as a tool to estimate cooling functions at various temperatures and densities.

With green lines we show the adopted approximation \eqref{{cool1}}. Obviously, our fit overestimates the cooling function at low densities and temperatures in excess of several hundred~K. However, in our model such high temperatures are only met at later evolutionary stages, when the density is quite high already. Red lines show the approximation presented by \cite{2001ApJ...557..736G}. This approximation is valid at moderate densities ($10^3-10^4$~cm$^{-3}$), but fails both at lower and higher densities.

\subsubsection{Stellar $L-M$ and $T_\text{eff}-M$ relations}

As soon as the central density reaches $10^{14}$~cm$^{-3}$ (the corresponding temperature exceeds 1000~K) we exclude an inner region from the numerical grid and introduce a sink cell with luminosity $L$ and effective temperature $T_\text{eff}$, which are associated with a central star we assume to be formed. During the further evolution the star accumulates mass via accretion from the rest of the molecular core. The $L$ and $T_\text{eff}$ evolve with the stellar mass as
\begin{align}
  &\lg L = -1.04 + 2.42 \left(\lg M + 1\right) \label{LM}\\
  &\lg T_\text{eff} = 3.49 + 0.4 \left(\lg M + 1\right). \label{TM}
\end{align}
These relations were obtained by fitting evolutionary tracks of accreting protostars presented by Yorke \& Bodenheimer (\cite{2008ASPC..387..189Y}, their Fig.~2).

\subsection{Head 2: Presta}

The second head of Gorynych is Presta, a one-dimensional model for calculating the chemical evolution of a spherically symmetric cloud irradiated by both diffuse external ultraviolet (UV) radiation and a central source (protostar) with assigned parameters. The modelled object is divided into concentric layered cells (the grid is independent of the Kamelung grid). The density, gas temperature, and dust temperature are specified individually for each cell. The cells do not interact and do not exchange matter. The UV radiation intensity is calculated from the dimensionless intensity values given at outer and inner boundaries of the computational domain, and from the column density ($A_{\rm V}$) corresponding to a given cell.

Presta is able to account for the evolution of an object by sequentially considering stages with different values of the parameters in each cell and at a given point in time. A typical size of the Kamelung grid is well in excess of 10,000 cells for calculations of collapse, and this is intractable for a chemical modelling. So, the Presta grid should be drastically shortened, to keep the Presta code operative. The transformation can be done in different ways, depending on the considered problem. In the presented research the chemical evolution is presented for 30 cells, which are selected uniformly in density for each considered time moment, with the physical parameters being interpolated from the Kamelung grid to the Presta grid.

Presta includes reactions with cosmic rays, two-particle reactions, photo-reactions (including those with photons induced by cosmic rays), dissociative recombination reactions with electrons and charged and neutral dust grains, adsorption, desorption, and chemical reactions on dust grains. While Presta is three-phase in its current implementation \cite{2022ARep...66..393B}, in this research we have limited ourselves with the two-phase model, taking into account only the processes occurring on the surfaces of dust grains and neglecting the presence of the mantle bulk. Next, we assume that fraction of the products of each surface reaction does not remain on the grain but immediately enters the gas phase, i.e. we take reactive desorption into account. In the current work, we take this fraction to be 5\%.

Kinetic equations are integrated individually in each cell up to a given age. Numerically they are integrated  using the VODE (Variable-coefficient Ordinary Differential Equation) solver\textbf{ \cite{byrne_thompson_vodef90web}.} Presta is capable of working with different sets of initial chemical composition and different chemical reaction grids, including those with deuterated compounds. The built-in chemical reaction rate analyser allows one to determine the main pathways of formation and destruction for selected chemical components, thus, assessing the influence of certain processes on the chemical evolution of an object. In this study, the ALCHEMIC chemical reaction network \cite{2011ApJS..196...25S} was used with some updates described in \cite{2019MNRAS.485.1843W}. In the output Presta provides both the radial distributions of the components at specified times and their column densities in the direction toward the center of the object.

The initial abundances of chemical elements relative to the number of hydrogen nuclei are listed in Table~\ref{inicomp}. These are so-called low-metal abundances from \cite{1996A&AS..119..111L}, which imply that refractory elements are depleted from the gas phase and are excluded from the chemical evolution.

\begin{table}
\caption{Initial abundances for the chemical model relative to the number of H nuclei.}
\label{inicomp}
\begin{tabular}{ll}
\hline
Species & Abundance \\
\hline
H                    & $10^{-5}$ \\
H$_2$                & 0.499995 \\
He                   & 0.09\\
C                  &  $7.30\cdot10^{-5}$\\
N                  &  $2.14\cdot10^{-5}$\\
O                  &  $1.76\cdot10^{-4}$\\
S                  &  $8.00\cdot10^{-8}$\\
Si                 &  $8.00\cdot10^{-9}$\\
Fe                 &  $3.00\cdot10^{-9}$\\
Mg                 &  $7.00\cdot10^{-9}$\\
Na                 &  $2.00\cdot10^{-9}$\\
\hline
\end{tabular}
\end{table}

\subsection{Head 3: Uran(IA)}

To calculate molecular line spectra, we adopt the Uran(IA) code described in detail in \cite{2004ARep...48..315P,2008ApJ...689..335P}. This code partly utilizes the scheme originally proposed and implemented in the publicly available one-dimensional code RATRAN \cite{2000A&A...362..697H}. The idea of the method is to solve the system of line radiative transfer equations with the accelerated $\Lambda$-iterations method, where the mean intensity $J_{\nu}$ is calculated using Monte Carlo ray sampling. All the molecular data needed for the line radiative transfer simulations were taken from the Leiden molecular database \cite{2005A&A...432..369S}.

\subsection{Setup for considered models}

For the study presented here we calculated dynamics, chemistry, and spectra for two star-forming objects, representative of low mass and high mass protostellar cores, which do not directly correspond to any specific object.

We begin our simulations by calculating the structure of the molecular core being in hydrostatic and thermal equilibrium. The core is exposed to the interstellar radiation field which is represented by blackbody spectrum with $T=2\times10^4$~K and dilution $D=10^{-16}$. The central H$_2$ concentration of the hydrostatic core is chosen to be $n_\text{hs}=10^4$~cm$^{-3}$, which is at the low limit for the observed starless cores~\cite{2017OAst...26..285D}. The outer boundary of the core $R_0$ is calculated as a result of the equilibrium solution and depends on the assumed core mass. The adopted stationary distributions can be considered as the generalization of the Bonnor-Ebert solution \cite{1956MNRAS.116..351B} where the density structure is obtained assuming the uniform temperature distribution. In our hydrostatic model, the temperature increases outwards because of the heating by interstellar radiation.

To initiate the collapse of the model cloud we double the density in each cell, and thus double the total mass of the core. As our experiments show, factor of two did not allow the core to find the new equilibrium state and forced it to collapse. In terms of Bonnor-Ebert interpretation, the possible hydrostatic state of the density-enhanced core turns out to be supercritical, i.e. unstable to collapse. Using this artificial approach we circumvent the problem of the very early molecular core evolution and do not address the physical mechanisms to initiate the collapse.  In reality, the collapse of the molecular cores may be initiated through the quasi-static contraction by ambipolar diffusion of magnetic field~\cite{2002ApJ...569..792L} or triggered by the influence of the external UV radiation~\cite{1997MNRAS.292..601S} or shocks~\cite{1976ApJ...207..484W}. We also note, that molecular cores could evolve dynamically even during their formation stages~\cite{2004RvMP...76..125M}, i.e. may not be considered as quasi-static objects at any stages. 

The problem of choosing the boundary conditions is coupled to the problem of initial conditions i.e. depends of the adopted physical paradigms which are still controversial. As a boundary condition, we assume zero velocity at the outer core border that fixes the position of the core boundary within the adopted Lagrangian formalism.

In our study, we consider the low-mass (LM) model with total mass $M_{\text{core}}=5~M_{\odot}$ and the high-mass (HM) model with $M_{\text{core}}=50~M_{\odot}$.
The initial distributions of density and temperature just before the hydrodynamical simulations are shown in Fig.~\ref{physcollapse}. Given the central density $n_0=2\times 10^4$ cm$^{-3}$, the free-fall time for both clouds is $t_\text{ff}=(3\pi/(32\,G\rho))^{1/2} = 0.26$~Myr.

The collapse of each core proceeds until the formation of the FHSC, which is then followed by the accretion phase. These stages will be described below.

\section{Results: Low mass versus high mass core evolution}

\subsection{Dynamics}

\begin{figure}
\centering
\includegraphics[width=0.49\textwidth]{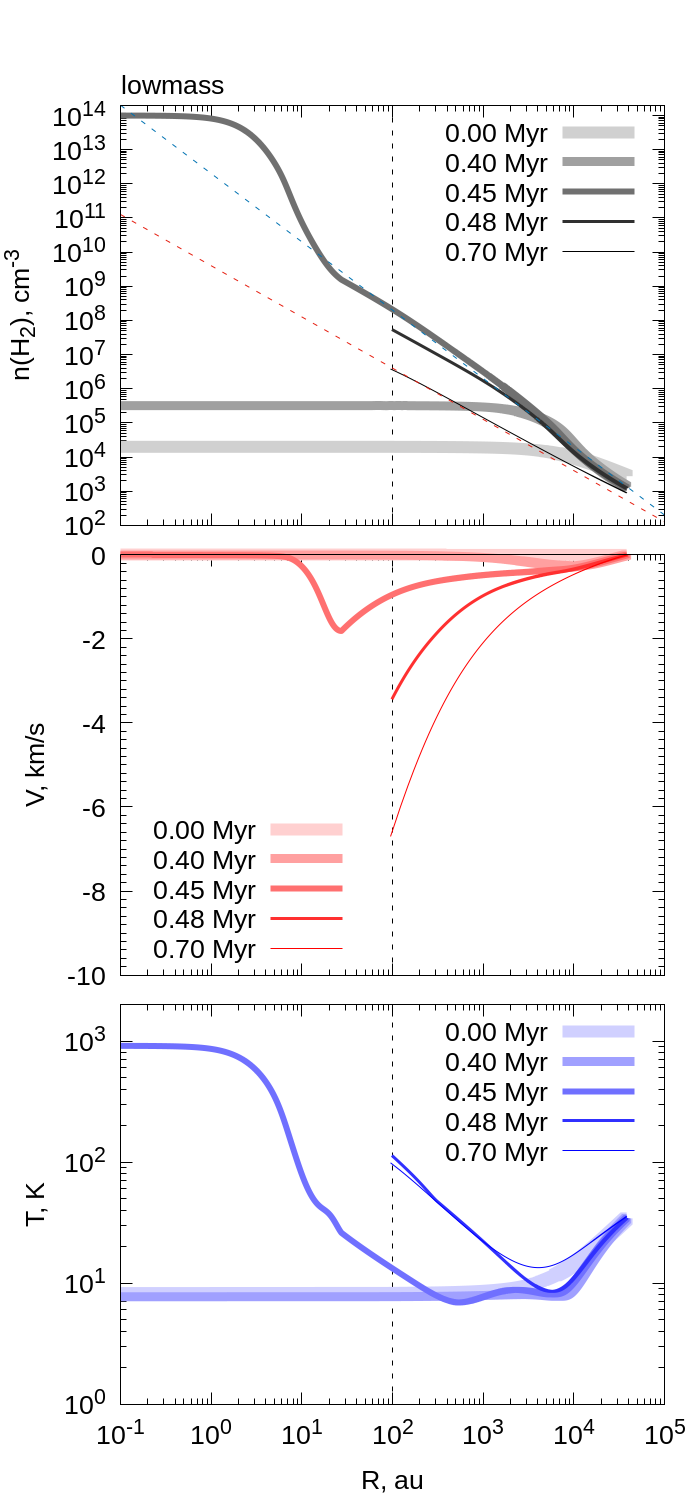}
\includegraphics[width=0.49\textwidth]{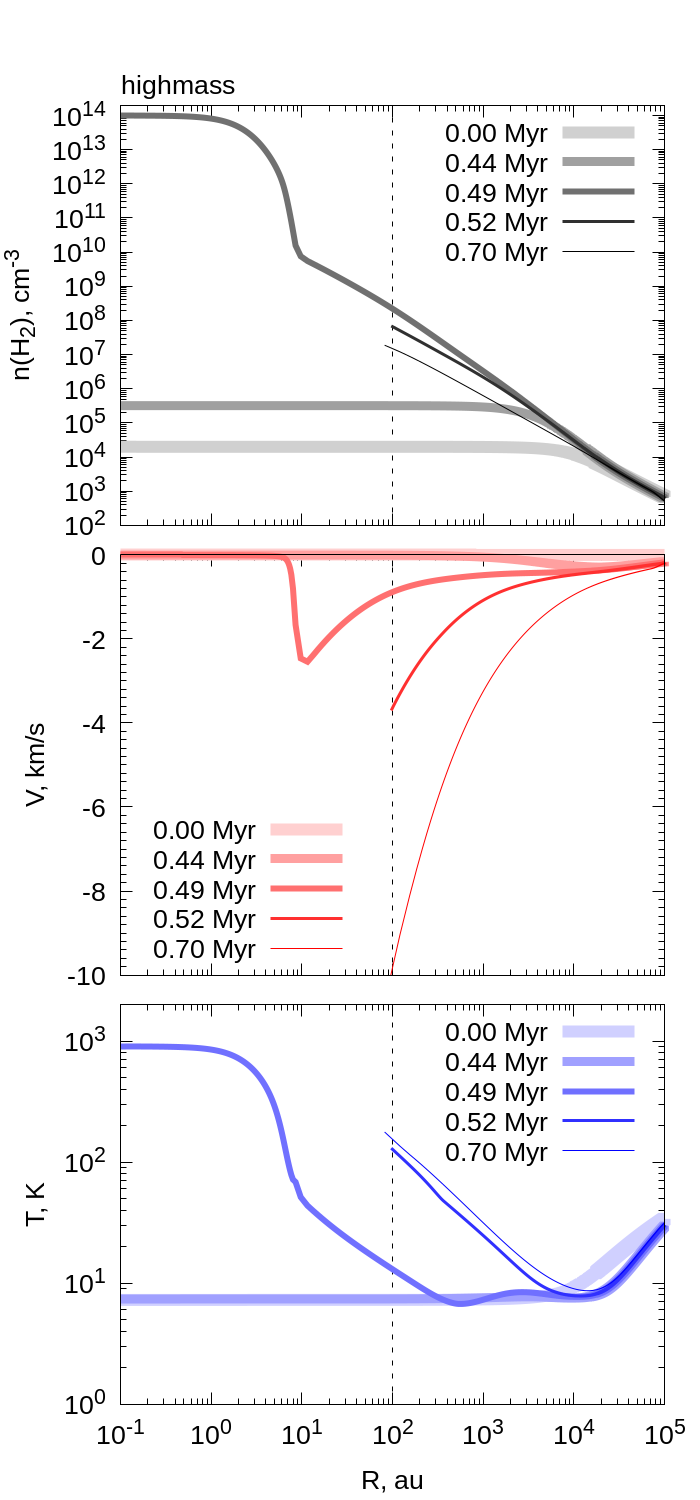}
\caption{Evolution of density, velocity and temperature distributions for the low mass (left) and high mass (right) cores. Blue and red lines in the left top panel represent the power laws $n\propto r^{-2}$ and $n\propto r^{-1.5}$, correspondingly.}
\label{physcollapse}
\end{figure}

The structures of the low mass and high mass model cores for the selected time moments are shown in right and left panels of Figure~\ref{physcollapse}, correspondingly. The process starts with nearly identical density distributions, albeit with different sizes for the cores of different masses (see the thickest lines on the top panels of Figure~\ref{physcollapse}). Given the same central densities for both cores, the difference in their masses is mostly related to the difference in their sizes. Save for the radial extent, initial temperature and density profiles are similar for both cores.

As the collapse proceeds, both cores gradually approach the state of the FHSC formation, which corresponds to the age of about 0.45 Myr for the low mass core and to the age of about 0.49 Myr for the high mass core, counting from the onset of the collapse. 
Note that these ages are about two times (1.73~t$_\text{ff}$ and 1.88~t$_\text{ff}$, correspondingly) higher than free fall time since the initial clouds were non-uniform so the collapse was not in free-fall mode.
By this time, about 4000 Lagrangian cells (of all 70000 cells allocated for this model) with total mass of 0.1~M$_\odot$, which were initially outside 100~au have moved inside this region. A density plateau forms in the very central region, where the future protostars are being formed. The velocities are nearly zero there, since the material stops moving, slowly growing denser as the outer envelope keeps falling onto it (Fig.~\ref{physcollapse}, middle panels).
The density outside the plateau follows the law $n\propto r^{-2}$ (shown with blue dashed line in the left top panel of Fig.~\ref{physcollapse}) which is the well known self-similar Larson-Penston solution for the isothermal collapse~\cite{1969MNRAS.144..425P,1969MNRAS.145..271L}.

As the density in the center rises to $10^{14}$~cm$^{-3}$, and the temperature grows to $10^3$~K, physics within the central region starts to involve processes beyond the Kamelung consideration, so when this density is reached in the innermost cell, the cells within the radius of 100 au are ``cut off'', and further on we follow the collapse of the envelope only.  The cutting value of 100 au is selected as this is the typical radius of protostellar disks around young stars. The disk will significantly affect the observational appearance both in the dust continuum and molecular lines but there is no way to account it correctly with the spherically symmetric approach in Kamelung. The material in the inner sink cell still gravitationally affects the envelope.

Thin lines in Fig.~\ref{physcollapse} represent the structure of the cores after the formation of the sink cell, and we consider this state as an early protostellar phase. Even being excluded from the computational domain, the protostars in Kamelung keep evolving and heating up, as the material of the shell falls onto them. At about 0.7 Myr the outer shell basically stops evolving, with the density slowly approaching the power law $n\propto r^{-1.5}$ profile, associated with accretion, and this is when we stop the computation. Simultaneously, the inner shell heats up, and the infall velocity raises. Like at the preceding evolutionary phases, the structure of the high mass core does not differ significantly from the structure of the low mass core. Again, the main difference is in the linear sizes of the cores, along with a minute difference in temperatures at later times.

The results of the collapse modelling are further passed to the Presta module for calculation of the chemical evolution.

\subsection{Chemistry}

\begin{figure}[b!]
\centering
\includegraphics[width=0.90\textwidth,clip=]{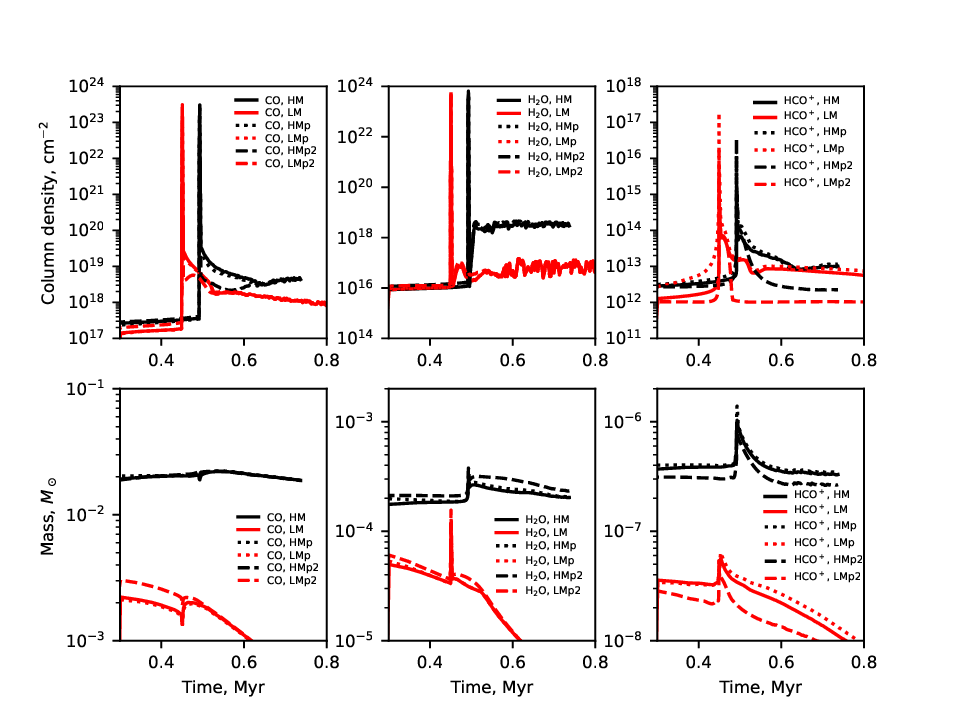}
\caption{Column densities and total masses of selected species as a function of time for the HM and LM models.}
\label{cd}
\end{figure}

Presta, the second ``head'' of Gorynych, is used to calculate the chemical evolution of the cores. As an input, we use the physical structure of the cores calculated at the previous step, but for a significantly reduced number of points. Figure~\ref{cd} shows the evolution of column densities and total masses for carbon monoxide, CO, as a tracer of molecular gas, water, H$_2$O, as one of the most important astrobiological species, and formyl ion, HCO$^+$, a typical ion, primarily found in molecular clouds and observed in a wide range of objects. 

As the collapse progresses, at a certain moment, which corresponds to the formation of the FHSC and transition from the pre-stellar phase to the protostellar phase, the column densities grow sharply for all the species. This transition heats up interior layers and evaporates grain mantles, as the temperature there rises up to few hundred K. The fall-off of the column density back to what it had been before signifies the onset of the accretion regime, when Kamelung algorithm removes the inner hot region (a Class~0 object) from further calculations, and we proceed with modeling only the infalling shell. Gradually, more and more material moves from the colder regions to the warmer ones, starting evaporate the dust grain mantles or at least altering the main chemical routes. The warmed-up region occupies small volume and does not make a significant impact on the core as a whole. This is visible on bottom panels of Figure~\ref{cd}, where we plot total masses of the selected molecules. Despite a very sharp increase of column densities in the direction of the core center, the total masses of the molecules do not change significantly if change at all.

As we mentioned before, we imply that the zero point for Presta is the same as the starting point for Kamelung, with the central density of the core being equal to $10^4$~cm$^{-3}$. This value is quite high to be an initial state of an actual molecular cloud. Since it is likely that the chemical evolution had started in a more tenuous environment and had lasted already for a while before the physical configuration has come to our ``zero point'', we need to evaluate the impact of this omitted phase on the chemical evolution. So we have also considered models with a pre-calculated molecular chemical composition, corresponding to a protracted initial stage of the core evolution. Specifically, we computed two models of the chemical evolution keeping an initial state of each core unchanged for $3\cdot10^5$ years. These models are designated LMp and HMp and are shown in Figure~\ref{cd} with dotted lines. In another two models, the initial chemical composition is computed for a uniform density of $10^2$~cm$^{-3}$, also for $3\cdot10^5$ years. These models are designated LMp2 and HMp2 and are shown in Figure~\ref{cd} with dashed lines. In all cases, the zero time moment corresponds to the onset of the collapse, so that time moments corresponding to the preliminary stage are negative.
 
As we see in Figure~\ref{cd}, taking into account of a preceding quiescent phase does not make much difference for CO and water. The dependence of computed column densities on the initial conditions is somewhat more noticeable for HCO$^+$ (top right panel in Figure~\ref{cd}), with $N($HCO$^+)$ being almost an order of magnitude smaller at later times in HMp2 and LMp2 models than in the other models, but this difference is mostly observed in the central part of the core, so that it has a less significant effect on the total mass of the ion and most probably will not be discernible in observations with finite beam size. A similar picture is observed in both cores for other molecules, including those not presented in figures. So, the main result the pre-calculated chemistry brought is the shift in the chemical age, adding extra 300 thousand years to it, but not altering the overall outline of the chemical evolution. This degeneracy makes speaking about the chemical age in terms of years meaningless, as it depends strongly on the definition of the starting point. More relevant is to refer to the age in terms of the collapse phases: pre-stellar phase, formation of the FHSC, and the accretion phase.

\begin{figure}[t!]
\centering
\includegraphics[width=\textwidth]{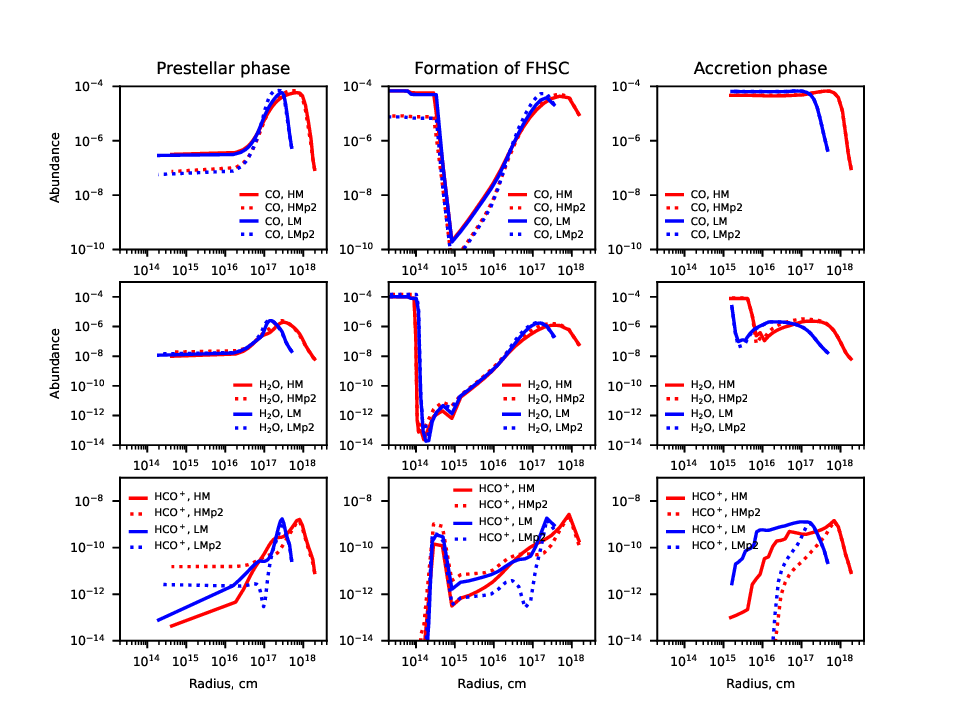}
\caption{Radial abundance profiles of the selected species at three time moments, representing the prestellar phase, the formation of the FHSC, and the protostellar phase.}
\label{radabun}
\end{figure}

Similar picture with minute variations is seen, when we compare radial distributions of the considered species for the two sets of models. On Figure~\ref{radabun} we present the radial abundance profiles corresponding to the mentioned stages. As we previously noted, most of the difference in the radial abundances at different stages is caused by the inner warm region, which is not just warm, but gradually heats up. At the prestellar stage both carbon monoxide and water demonstrate higher gas-phase abundances.

At the second stage we see a sharp increase of CO and water abundances in the center of the core caused by the evaporation of icy mantles. Still most of the core mass resides in the cold relatively unevolved region of the shell and does not get impacted by the interior processes. We do see a significant differences in HCO$^+$ abundances between the models with precomputed low-density chemistry (HMp2 and LMp2) and the models with atomic initial composition (HM and LM). However, as we said before, these differences are most significant in the central region, which only comprises a minor fraction of the core volume.

Comparing the evolution of the chemical composition of the cores of different masses reveals some differences, worth investigating. The radial profile of the CO abundance rises as the temperature grows with the formation of the protostar in both models, though for the LM core the growth is more prominent, up to four orders of magnitude in the innermost region, while for the HM core the growth does not exceed a factor of two. Same trend is observed for the evolution of HCO$^+$, as we see in Fig.~\ref{radabun}. The radial distribution of water during the transition evolutionary stage also expresses an interesting feature, as for the central regions of the cores, water is more abundant in the low mass model than in the high mass one. Though for more progressed protostellar stage the situation changes back, as the abundance of water in the high mass model gets higher again. The differences observed in Fig.~\ref{radabun} for the HM and LM plots are explained by the scale factor. The more massive cloud gain its mass due to the larger size. Minding that, one can say that the general picture of the evolution of radial abundance for water is the same for both the cores.

The specifics of chemistry under the conditions of the ongoing collapse, and explaining its peculiarities, with differences and similarities in different models, is a challenging problem on its own and is an objective of the separate further study.

\subsection{Line profiles and integral intensity}

\begin{figure}
\centering
\includegraphics[width=0.49\textwidth]{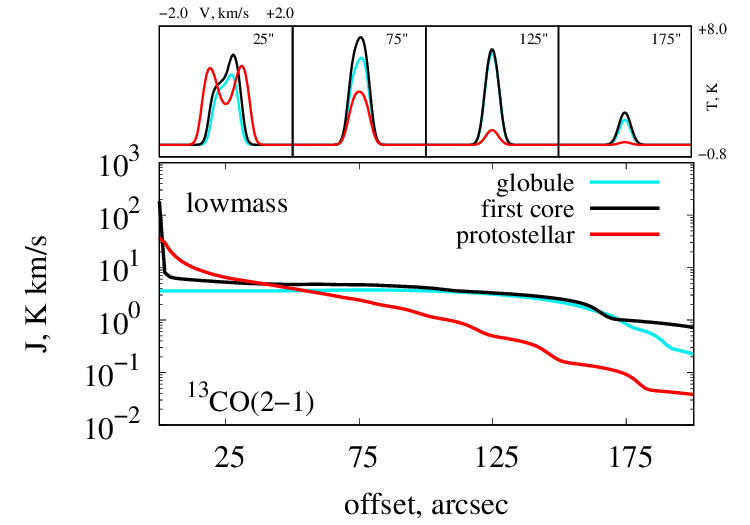}
\includegraphics[width=0.49\textwidth]{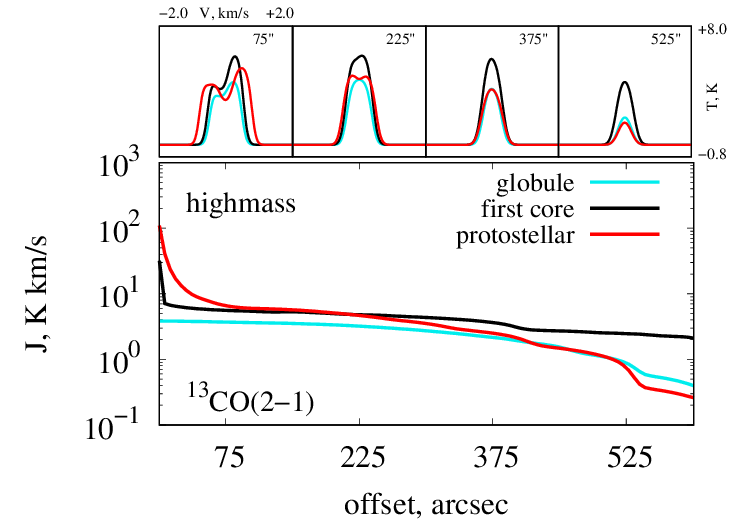} \\
\includegraphics[width=0.49\textwidth,clip=]{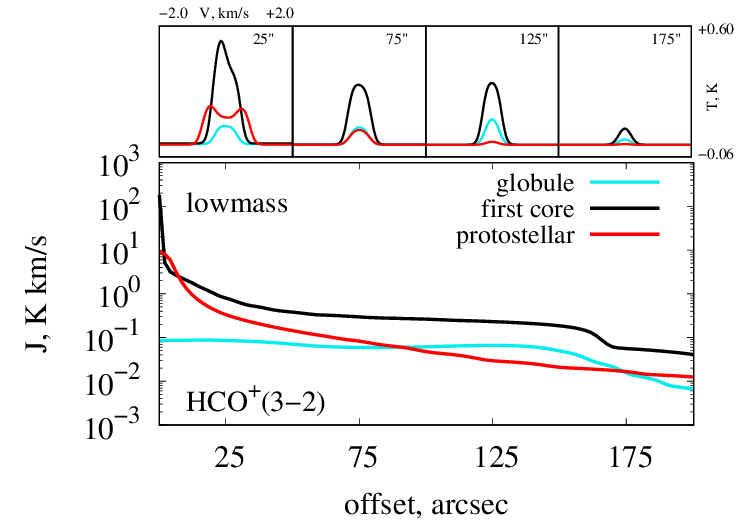}
\includegraphics[width=0.49\textwidth,clip=]{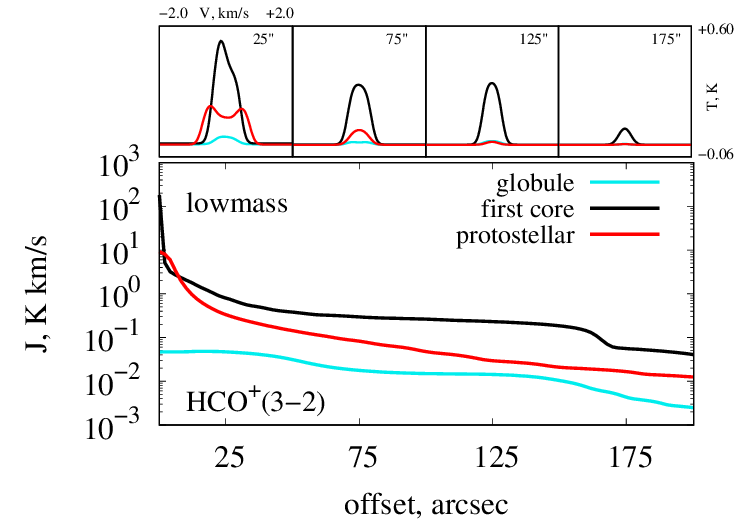} \\
\includegraphics[width=0.49\textwidth]{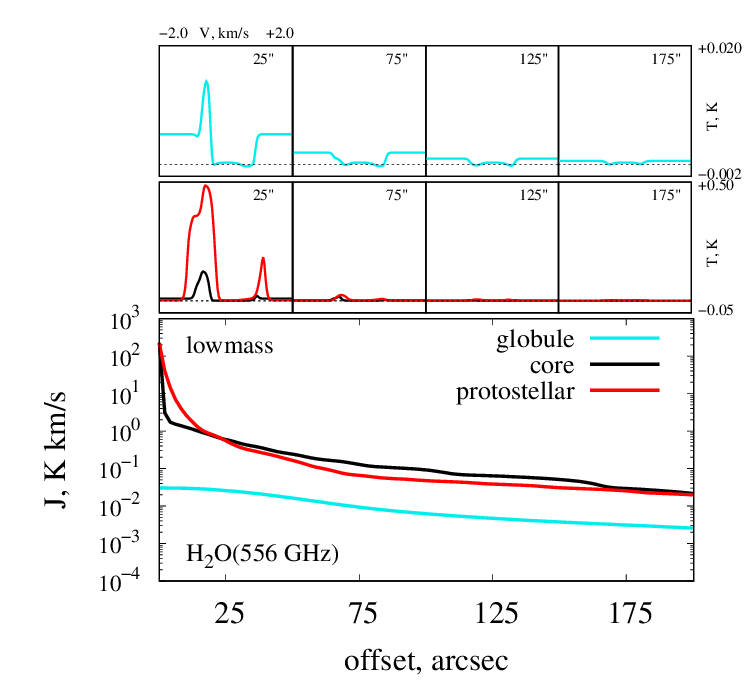}
\includegraphics[width=0.49\textwidth]{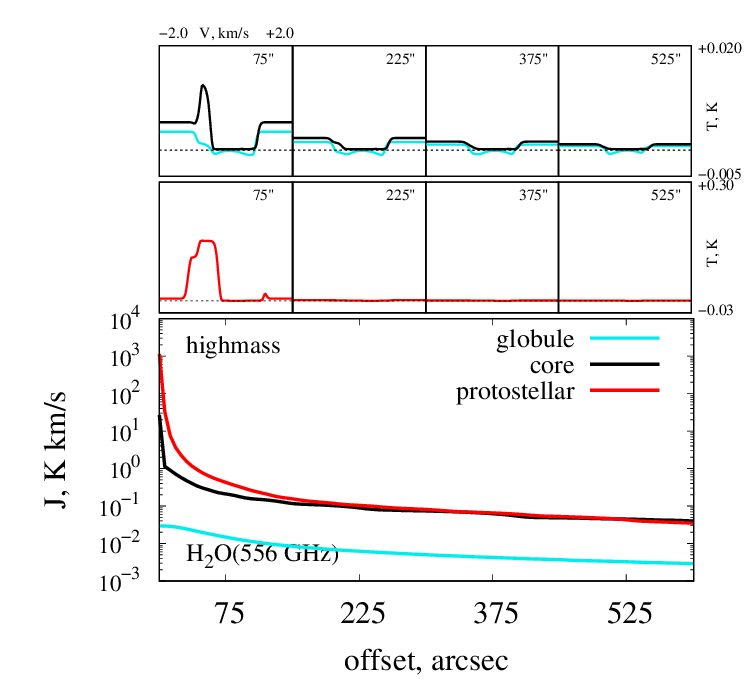}
\caption{Spectra and integral intensity distributions for $^{13}$CO(2--1) (top), HCO$^+$(3--2) (middle), and H$_2$O(110--101) (bottom) emission corresponding to three stages of evolution: the prestellar core (globlule) ($n_0($H$_2)=2 \times 10^5$ cm$^{-3}$, $t\sim 40$ thousand years for LM, $t\sim 44$ thousand years for HM), the FHSC (core) ($n_0($H$_2)=10^{14}$ cm$^{-3}$, $t\sim 45$ thousand years for LM, and $t\sim 49$ thousand years for HM), and the protostellar object (protostar) ($t\sim700$ thousand years)}
\label{spectra}
\end{figure}

In Figure~\ref{spectra}, we show the spatial distributions of the integral intensity $J$ [K\,km\,s$^{-1}$] of the $^{13}$CO(2--1), HCO$^+$(3--2), and H$_2$O(110--101) lines for the low-mass and high-mass models. We selected three time moments corresponding to the prestellar core ($n_0($H$_2)=2 \times 10^5$ cm$^{-3}$, $t\sim 40$ thousand years for LM, $t\sim 44$ thousand years for HM), the FHSC ($n_0($H$_2)=10^{14}$ cm$^{-3}$, $t\sim 45$ thousand years for LM, and $t\sim 49$ thousand years for HM), and the protostellar object ($t\sim700$ thousand years) for the HM and LM models. On the same panels, we also show the line profiles towards four positions along the core (see offsets from the core center in the upper corners of the plots). No beam convolution is performed.

We first describe the results for the LM model. The $J$-distribution of $^{13}$CO(2--1) for the prestellar core is nearly flat inside 125~arcsec due to significant CO depletion toward the center. At the FHSC stage, the $J$-intensity  stays nearly the same as in the prestellar phase except the most inner part where it has a strong maximum toward the center of the core (due to the high CO abundance and temperature in the vicinity of the FHSC). At the protostelar phase, the $J$-distribution monotonically decreases from the center of the core (due to exhaustion of the envelope and the negative temperature gradient).
The line profiles toward 25~arcsec position are red-skewed for the prestellar and FHSC stages, and double-peaked for protostellar phase.
The asymmetry of the red-skewed profile is caused by the interplay between the positive temperature gradient in the outer parts of envelope (due to external heating) and the infall velocity.   

The morphology of HCO$^+$(3--2) integral intensity distributions are similar to those of $^{13}$CO(2--1). The noticeable difference is that the prestellar distribution is significantly lower than the FHSC distribution over the entire core extent. We also note that spectrum toward 25~arcsec position at the FHSC stage is blue-skewed, unlike $^{13}$CO(2--1).

The H$_2$O emission is sensitive to high density and temperature. This is the reason why H$_2$O(110--101) line is very week at the prestellar core phase. In fact, this line manifests itself as absorption with respect to the dust continuum emission. The distributions of integral intensity for the FHSC and the protostellar core are centrally peaked. The line profiles toward 25~arcsec are highly asymmetric with profound central self-absorption dips.

The most significant difference between the HM and LM cases is that the spatial extent of line emission is about three times larger with the same morphological features as in the low-mass case.

\section{Discussion and Conclusions}

In this paper we present a complex model of a collapsing protostellar core, which combines the hydrodynamical part to simulate the evolution of the core physical structure, the chemical part to compute time-dependent molecular abundances in a post-processing mode, and the radiative part to convert the physical structure and the computed abundances into synthetic line profiles. This approach is by no means new and has been invoked in various forms by many researchers. Our specific goal is to consider advanced evolutionary stages of the collapsing core at the expense of simplified geometry. This allows us to follow the core evolution to the formation of the FHSC and beyond, into the accretion phase. We cannot treat the physics in the immediate vicinity of the forming protostar, which limits our options, however, the neglected region is quite small and will hardly influence the observational appearance of the core in lines of the considered molecules. Of course, this may not be true for some species, which form exclusively in the vicinity of the protostar. But those species that form in the envelope are treated adequately in Gorynych. This makes it a powerful tool to consider various observational tracers of the evolutionary stages around the formation of the FHSC and later.

Another limitation of our model is that we cannot treat the core rotation and its magnetic field. Both factors can and do lead to the core anisotropy, which defies our assumption of its spherical symmetry, and may also alter the core dynamical behavior. Thus, our model should be considered as a tool to study in detail various chemical factors with reasonably realistic evolution of density and temperature rather than an accurate representation of a dynamical evolution of a collapsing protostar.

Notably, the innermost regions of our collapsing clouds, both high-mass and low-mass, are very similar in their initial conditions. Given fixed central densities, their density and temperature distributions are nearly identical inside $10^4$~au. The difference appears in the outer parts of the clouds but the morphology of the distributions are similar. On the other hand, the advanced stage of collapse for both cores is likely to proceed in a self-consistent regime following Larson-Penston solution\textbf{~\cite{1969MNRAS.144..425P,1969MNRAS.145..271L}}. This is the reason why the dynamical evolution of our low-mass and high-mass clouds are very similar up to the late accretion phases where the difference in the available mass will play a role.

As we start our dynamical modelling at a rather high density, which implies some preceding chemical evolution. To elucidate its possible effect, we add a quiescent phase to allow some molecules to attain non-zero abundances at the beginning of the collapse. This phase does not seem to make a difference for water and CO, as their abundances quickly adapt to changing environmental conditions. Some variations are seen in the central HCO$^+$ abundances, which makes this ion a promising tracer of the initial conditions for collapsing cores.

\section{Acknowledgements}

The authors are deeply grateful to the anonymous reviewer for their careful reading of the manuscript, constructive criticism, and valuable suggestions, which have significantly improved the quality of this work.

\bibliographystyle{inasan}
\bibliography{Gorynych}

@ARTICLE{2005A&A...432..369S,
       author = {{Sch{\"o}ier}, F.~L. and {van der Tak}, F.~F.~S. and {van Dishoeck}, E.~F. and {Black}, J.~H.},
        title = "{An atomic and molecular database for analysis of submillimetre line observations}",
      journal = {\aap},
         year = 2005,
        month = mar,
       volume = {432},
       number = {1},
        pages = {369-379},
          doi = {10.1051/0004-6361:20041729},
archivePrefix = {arXiv},
       eprint = {astro-ph/0411110},
 primaryClass = {astro-ph},
       adsurl = {https://ui.adsabs.harvard.edu/abs/2005A&A...432..369S}
}

@ARTICLE{2000A&A...362..697H,
       author = {{Hogerheijde}, M.~R. and {van der Tak}, F.~F.~S.},
        title = "{An accelerated Monte Carlo method to solve two-dimensional radiative transfer and molecular excitation. With applications to axisymmetric models of star formation}",
      journal = {\aap},
         year = 2000,
        month = oct,
       volume = {362},
        pages = {697-710},
          doi = {10.48550/arXiv.astro-ph/0008169},
archivePrefix = {arXiv},
       eprint = {astro-ph/0008169},
 primaryClass = {astro-ph},
       adsurl = {https://ui.adsabs.harvard.edu/abs/2000A&A...362..697H}
}

@ARTICLE{2008ApJ...689..335P,
       author = {{Pavlyuchenkov}, Ya. and {Wiebe}, D. and {Shustov}, B. and {Henning}, Th. and {Launhardt}, R. and {Semenov}, D.},
        title = "{Molecular Emission Line Formation in Prestellar Cores}",
      journal = {\apj},
         year = 2008,
        month = dec,
       volume = {689},
       number = {1},
        pages = {335-350},
          doi = {10.1086/592564},
archivePrefix = {arXiv},
       eprint = {0808.2375},
 primaryClass = {astro-ph},
       adsurl = {https://ui.adsabs.harvard.edu/abs/2008ApJ...689..335P}
}

@ARTICLE{2004ARep...48..315P,
       author = {{Pavlyuchenkov}, Ya. N. and {Shustov}, B.~M.},
        title = "{A Method for Molecular-Line Radiative-Transfer Computations and Its Application to a Two-Dimensional Model for the Starless Core L1544}",
      journal = {Astronomy Reports},
         year = 2004,
        month = apr,
       volume = {48},
       number = {4},
        pages = {315-326},
          doi = {10.1134/1.1704676},
       adsurl = {https://ui.adsabs.harvard.edu/abs/2004ARep...48..315P}
}

@ARTICLE{1996A&AS..119..111L,
       author = {{Lee}, H. -H. and {Bettens}, R.~P.~A. and {Herbst}, E.},
        title = "{Fractional abundances of molecules in dense interstellar clouds: A compendium of recent model results.}",
      journal = {\aaps},
         year = 1996,
        month = oct,
       volume = {119},
        pages = {111-114},
       adsurl = {https://ui.adsabs.harvard.edu/abs/1996A&AS..119..111L}
}

@ARTICLE{2019MNRAS.485.1843W,
       author = {{Wiebe}, Dmitri S. and {Molyarova}, Tamara S. and {Akimkin}, Vitaly V. and {Vorobyov}, Eduard I. and {Semenov}, Dmitry A.},
        title = "{Luminosity outburst chemistry in protoplanetary discs: going beyond standard tracers}",
      journal = {\mnras},
         year = 2019,
        month = may,
       volume = {485},
       number = {2},
        pages = {1843-1863},
          doi = {10.1093/mnras/stz512},
archivePrefix = {arXiv},
       eprint = {1902.07475},
 primaryClass = {astro-ph.EP},
       adsurl = {https://ui.adsabs.harvard.edu/abs/2019MNRAS.485.1843W}
}

@ARTICLE{2011ApJS..196...25S,
       author = {{Semenov}, D. and {Wiebe}, D.},
        title = "{Chemical Evolution of Turbulent Protoplanetary Disks and the Solar Nebula}",
      journal = {\apjs},
         year = 2011,
        month = oct,
       volume = {196},
       number = {2},
          eid = {25},
        pages = {25},
          doi = {10.1088/0067-0049/196/2/25},
archivePrefix = {arXiv},
       eprint = {1104.4358},
 primaryClass = {astro-ph.GA},
       adsurl = {https://ui.adsabs.harvard.edu/abs/2011ApJS..196...25S}
}

@ARTICLE{2022ARep...66..393B,
       author = {{Borshcheva}, E.~V. and {Wiebe}, D.~S.},
        title = "{Luminosity Outbursts in Protoplanetary Discs: A Three-Phase Astrochemical Model}",
      journal = {Astronomy Reports},
         year = 2022,
        month = may,
       volume = {66},
       number = {5},
        pages = {393-418},
          doi = {10.1134/S1063772922060014},
       adsurl = {https://ui.adsabs.harvard.edu/abs/2022ARep...66..393B}
}

@INPROCEEDINGS{2008ASPC..387..189Y,
       author = {{Yorke}, H.~W. and {Bodenheimer}, P.},
        title = "{Theoretical Developments in Understanding Massive Star Formation}",
    booktitle = {Massive Star Formation: Observations Confront Theory},
         year = 2008,
       editor = {{Beuther}, H. and {Linz}, H. and {Henning}, Th.},
       series = {Astronomical Society of the Pacific Conference Series},
       volume = {387},
        month = may,
        pages = {189},
       adsurl = {https://ui.adsabs.harvard.edu/abs/2008ASPC..387..189Y}
}

@ARTICLE{2009ARep...53..611K,
       author = {{Kirsanova}, M.~S. and {Wiebe}, D.~S. and {Sobolev}, A.~M.},
        title = "{Chemodynamical evolution of gas near an expanding HII region}",
      journal = {Astronomy Reports},
         year = 2009,
        month = jul,
       volume = {53},
       number = {7},
        pages = {611-633},
          doi = {10.1134/S106377290907004X},
       adsurl = {https://ui.adsabs.harvard.edu/abs/2009ARep...53..611K}
}

@ARTICLE{2001ApJ...557..736G,
       author = {{Goldsmith}, Paul F.},
        title = "{Molecular Depletion and Thermal Balance in Dark Cloud Cores}",
      journal = {\apj},
         year = 2001,
        month = aug,
       volume = {557},
       number = {2},
        pages = {736-746},
          doi = {10.1086/322255},
       adsurl = {https://ui.adsabs.harvard.edu/abs/2001ApJ...557..736G}
}

@ARTICLE{2023RMxAA..59..327C,
       author = {{Chatzikos}, M. and {Bianchi}, S. and {Camilloni}, F. and {Chakraborty}, P. and {Gunasekera}, C.~M. and {Guzm{\'a}n}, F. and {Milby}, J.~S. and {Sarkar}, A. and {Shaw}, G. and {van Hoof}, P.~A.~M. and {Ferland}, G.~J.},
        title = "{The 2023 Release of Cloudy}",
      journal = {\rmxaa},
         year = 2023,
        month = oct,
       volume = {59},
        pages = {327-343},
          doi = {10.22201/ia.01851101p.2023.59.02.12},
archivePrefix = {arXiv},
       eprint = {2308.06396},
 primaryClass = {astro-ph.GA},
       adsurl = {https://ui.adsabs.harvard.edu/abs/2023RMxAA..59..327C}
}

@ARTICLE{2023RNAAS...7..246G,
       author = {{Gunasekera}, Chamani M. and {van Hoof}, Peter A.~M. and {Chatzikos}, Marios and {Ferland}, Gary J.},
        title = "{The 23.01 Release of Cloudy}",
      journal = {Research Notes of the American Astronomical Society},
         year = 2023,
        month = nov,
       volume = {7},
       number = {11},
          eid = {246},
        pages = {246},
          doi = {10.3847/2515-5172/ad0e75},
archivePrefix = {arXiv},
       eprint = {2311.10163},
 primaryClass = {astro-ph.GA},
       adsurl = {https://ui.adsabs.harvard.edu/abs/2023RNAAS...7..246G}
}

@ARTICLE{2006ApJS..164..506L,
       author = {{Le Petit}, Franck and {Nehm{\'e}}, Cyrine and {Le Bourlot}, Jacques and {Roueff}, Evelyne},
        title = "{A Model for Atomic and Molecular Interstellar Gas: The Meudon PDR Code}",
      journal = {\apjs},
         year = 2006,
        month = jun,
       volume = {164},
       number = {2},
        pages = {506-529},
          doi = {10.1086/503252},
archivePrefix = {arXiv},
       eprint = {astro-ph/0602150},
 primaryClass = {astro-ph},
       adsurl = {https://ui.adsabs.harvard.edu/abs/2006ApJS..164..506L}
}

@ARTICLE{2020INASR...5..294K,
       author = {{Kochina}, O.~V. and {Van Looveren}, G.},
        title = "{PRESTALINE: a synthetic spectra modeling tool for star forming regions}",
      journal = {INASAN Science Reports},
         year = 2020,
        month = nov,
       volume = {5},
        pages = {294-297},
          doi = {10.26087/INASAN.2020.5.5.016},
       adsurl = {https://ui.adsabs.harvard.edu/abs/2020INASR...5..294K}
}

@ARTICLE{2022AstL...48..517L,
       author = {{Looveren}, G. Van and {Kochina}, O.~V. and {Wiebe}, D.~S. and {Buslaeva}, A.~I.},
        title = "{DR21(OH) SUB-CORES: Inferring an Evolutionary Status Using the Prestaline Tool*}",
      journal = {Astronomy Letters},
         year = 2022,
        month = sep,
       volume = {48},
       number = {9},
        pages = {517-529},
          doi = {10.1134/S1063773722090043},
       adsurl = {https://ui.adsabs.harvard.edu/abs/2022AstL...48..517L}
}

@ARTICLE{2007A&A...468..627V,
       author = {{van der Tak}, F.~F.~S. and {Black}, J.~H. and {Sch{\"o}ier}, F.~L. and {Jansen}, D.~J. and {van Dishoeck}, E.~F.},
        title = "{A computer program for fast non-LTE analysis of interstellar line spectra. With diagnostic plots to interpret observed line intensity ratios}",
      journal = {\aap},
         year = 2007,
        month = jun,
       volume = {468},
       number = {2},
        pages = {627-635},
          doi = {10.1051/0004-6361:20066820},
archivePrefix = {arXiv},
       eprint = {0704.0155},
 primaryClass = {astro-ph},
       adsurl = {https://ui.adsabs.harvard.edu/abs/2007A&A...468..627V}
}

@ARTICLE{2013ARep...57..818K,
       author = {{Kochina}, O.~V. and {Wiebe}, D.~S. and {Kalenskii}, S.~V. and {Vasyunin}, A.~I.},
        title = "{Modeling of the formation of complex molecules in protostellar objects}",
      journal = {Astronomy Reports},
         year = 2013,
        month = nov,
       volume = {57},
       number = {11},
        pages = {818-832},
          doi = {10.1134/S1063772913110036},
       adsurl = {https://ui.adsabs.harvard.edu/abs/2013ARep...57..818K}
}

@ARTICLE{2020ApJS..249...26J,
       author = {{Jin}, Mihwa and {Garrod}, Robin T.},
        title = "{Formation of Complex Organic Molecules in Cold Interstellar Environments through Nondiffusive Grain-surface and Ice-mantle Chemistry}",
      journal = {\apjs},
         year = 2020,
        month = aug,
       volume = {249},
       number = {2},
          eid = {26},
        pages = {26},
          doi = {10.3847/1538-4365/ab9ec8},
archivePrefix = {arXiv},
       eprint = {2006.11127},
 primaryClass = {astro-ph.GA},
       adsurl = {https://ui.adsabs.harvard.edu/abs/2020ApJS..249...26J}
}

@ARTICLE{2022ApJ...935..133J,
       author = {{Jin}, Miwha and {Lam}, Ka Ho and {McClure}, Melissa K. and {van Scheltinga}, Jeroen Terwisscha and {Li}, Zhi-Yun and {Boogert}, Adwin and {Herbst}, Eric and {Davis}, Shane W. and {Garrod}, Robin T.},
        title = "{Ice Age: Chemodynamical Modeling of Cha-MMS1 to Predict New Solid-phase Species for Detection with JWST}",
      journal = {\apj},
         year = 2022,
        month = aug,
       volume = {935},
       number = {2},
          eid = {133},
        pages = {133},
          doi = {10.3847/1538-4357/ac8006},
archivePrefix = {arXiv},
       eprint = {2207.04269},
 primaryClass = {astro-ph.SR},
       adsurl = {https://ui.adsabs.harvard.edu/abs/2022ApJ...935..133J}
}

@ARTICLE{2012ApJ...758...86F,
       author = {{Furuya}, Kenji and {Aikawa}, Yuri and {Tomida}, Kengo and {Matsumoto}, Tomoaki and {Saigo}, Kazuya and {Tomisaka}, Kohji and {Hersant}, Franck and {Wakelam}, Valentine},
        title = "{Chemistry in the First Hydrostatic Core Stage by Adopting Three-dimensional Radiation Hydrodynamic Simulations}",
      journal = {\apj},
         year = 2012,
        month = oct,
       volume = {758},
       number = {2},
          eid = {86},
        pages = {86},
          doi = {10.1088/0004-637X/758/2/86},
archivePrefix = {arXiv},
       eprint = {1207.6693},
 primaryClass = {astro-ph.SR},
       adsurl = {https://ui.adsabs.harvard.edu/abs/2012ApJ...758...86F}
}

@ARTICLE{2016ApJ...822...12H,
       author = {{Hincelin}, U. and {Commer{\c{c}}on}, B. and {Wakelam}, V. and {Hersant}, F. and {Guilloteau}, S. and {Herbst}, E.},
        title = "{Chemical and Physical Characterization of Collapsing Low-mass Prestellar Dense Cores}",
      journal = {\apj},
         year = 2016,
        month = may,
       volume = {822},
       number = {1},
          eid = {12},
        pages = {12},
          doi = {10.3847/0004-637X/822/1/12},
archivePrefix = {arXiv},
       eprint = {1603.02529},
 primaryClass = {astro-ph.GA},
       adsurl = {https://ui.adsabs.harvard.edu/abs/2016ApJ...822...12H}
}

@ARTICLE{2013ApJ...775...44H,
       author = {{Hincelin}, U. and {Wakelam}, V. and {Commer{\c{c}}on}, B. and {Hersant}, F. and {Guilloteau}, S.},
        title = "{Survival of Interstellar Molecules to Prestellar Dense Core Collapse and Early Phases of Disk Formation}",
      journal = {\apj},
         year = 2013,
        month = sep,
       volume = {775},
       number = {1},
          eid = {44},
        pages = {44},
          doi = {10.1088/0004-637X/775/1/44},
archivePrefix = {arXiv},
       eprint = {1307.6868},
 primaryClass = {astro-ph.SR},
       adsurl = {https://ui.adsabs.harvard.edu/abs/2013ApJ...775...44H}
}

@ARTICLE{2020A&A...643A.108C,
       author = {{Coutens}, A. and {Commer{\c{c}}on}, B. and {Wakelam}, V.},
        title = "{Chemical evolution during the formation of a protoplanetary disk}",
      journal = {\aap},
         year = 2020,
        month = nov,
       volume = {643},
          eid = {A108},
        pages = {A108},
          doi = {10.1051/0004-6361/202038437},
archivePrefix = {arXiv},
       eprint = {2010.05108},
 primaryClass = {astro-ph.SR},
       adsurl = {https://ui.adsabs.harvard.edu/abs/2020A&A...643A.108C}
}

@ARTICLE{2018A&A...615A..15S,
       author = {{Sipil{\"a}}, O. and {Caselli}, P.},
        title = "{Hydrodynamics with gas-grain chemistry and radiative transfer: comparing dynamical and static models}",
      journal = {\aap},
         year = 2018,
        month = jul,
       volume = {615},
          eid = {A15},
        pages = {A15},
          doi = {10.1051/0004-6361/201732326},
archivePrefix = {arXiv},
       eprint = {1801.08004},
 primaryClass = {astro-ph.GA},
       adsurl = {https://ui.adsabs.harvard.edu/abs/2018A&A...615A..15S}
}

@ARTICLE{2016A&A...587A..60F,
       author = {{Frimann}, S. and {J{\o}rgensen}, J.~K. and {Padoan}, P. and {Haugb{\o}lle}, T.},
        title = "{Protostellar accretion traced with chemistry. Comparing synthetic C$^{18}$O maps of embedded protostars to real observations}",
      journal = {\aap},
         year = 2016,
        month = feb,
       volume = {587},
          eid = {A60},
        pages = {A60},
          doi = {10.1051/0004-6361/201527622},
archivePrefix = {arXiv},
       eprint = {1512.00416},
 primaryClass = {astro-ph.SR},
       adsurl = {https://ui.adsabs.harvard.edu/abs/2016A&A...587A..60F}
}

@ARTICLE{2019MNRAS.486.2525K,
       author = {{Kirsanova}, Maria S. and {Wiebe}, Dmitri S.},
        title = "{Merged H/H$_{2}$ and C$^{+}$/C/CO transitions in the Orion Bar}",
      journal = {\mnras},
         year = 2019,
        month = jun,
       volume = {486},
       number = {2},
        pages = {2525-2534},
          doi = {10.1093/mnras/stz983},
archivePrefix = {arXiv},
       eprint = {1904.04423},
 primaryClass = {astro-ph.GA},
       adsurl = {https://ui.adsabs.harvard.edu/abs/2019MNRAS.486.2525K}
}

@ARTICLE{2003ARep...47..176P,
       author = {{Pavlyuchenkov}, Ya. N. and {Shustov}, B.~M. and {Shematovich}, V.~I. and {Wiebe}, D.~S. and {Li}, Zhi-Yun},
        title = "{A Coupled, Dynamical and Chemical Model for the Prestellar Core L1544: Comparison of Modeled and Observed C18O, HCO+, and CS Emission Spectra}",
      journal = {Astronomy Reports},
         year = 2003,
        month = mar,
       volume = {47},
       number = {3},
        pages = {176-185},
          doi = {10.1134/1.1562212},
       adsurl = {https://ui.adsabs.harvard.edu/abs/2003ARep...47..176P}
}

@ARTICLE{1997MNRAS.292..601S,
       author = {{Shematovich}, V.~I. and {Shustov}, B.~M. and {Wiebe}, D.~S.},
        title = "{Self-consistent model of chemical and dynamical evolution of protostellar clouds}",
      journal = {\mnras},
         year = 1997,
        month = dec,
       volume = {292},
       number = {3},
        pages = {601-610},
          doi = {10.1093/mnras/292.3.601},
       adsurl = {https://ui.adsabs.harvard.edu/abs/1997MNRAS.292..601S}
}

@ARTICLE{2013ARep...57..641P,
       author = {{Pavlyuchenkov}, Ya. N. and {Zhilkin}, A.~G.},
        title = "{A multicomponent model for computing the thermal structure of collapsing protostellar clouds}",
      journal = {Astronomy Reports},
         year = 2013,
        month = sep,
       volume = {57},
       number = {9},
        pages = {641-656},
          doi = {10.1134/S1063772913090035},
       adsurl = {https://ui.adsabs.harvard.edu/abs/2013ARep...57..641P}
}

@ARTICLE{2015ARep...59..133P,
       author = {{Pavlyuchenkov}, Ya. N. and {Zhilkin}, A.~G. and {Vorobyov}, E.~I. and {Fateeva}, A.~M.},
        title = "{The thermal structure of a protostellar envelope}",
      journal = {Astronomy Reports},
         year = 2015,
        month = feb,
       volume = {59},
       number = {2},
        pages = {133-144},
          doi = {10.1134/S1063772915020067},
archivePrefix = {arXiv},
       eprint = {1502.04835},
 primaryClass = {astro-ph.GA},
       adsurl = {https://ui.adsabs.harvard.edu/abs/2015ARep...59..133P}
}

@misc{byrne_thompson_vodef90web,
  author       = {Byrne, G. D. and Thompson, S.},
  title        = {{VODE} f90 {ODE} Solver Web Support Page},
  howpublished = {Website},
  year         = {2008},
  url          = {http://www.radford.edu/~thompson/vodef90web/index.html},
  note         = {Online; accessed INSERT-ACCESS-DATE-HERE}
}

@ARTICLE{1956MNRAS.116..351B,
       author = {{Bonnor}, W.~B.},
        title = "{Boyle's Law and gravitational instability}",
      journal = {\mnras},
         year = 1956,
        month = jan,
       volume = {116},
        pages = {351},
          doi = {10.1093/mnras/116.3.351},
       adsurl = {https://ui.adsabs.harvard.edu/abs/1956MNRAS.116..351B}
}

@ARTICLE{2002ApJ...569..792L,
       author = {{Li}, Zhi-Yun and {Shematovich}, V.~I. and {Wiebe}, D.~S. and {Shustov}, B.~M.},
        title = "{A Coupled Dynamical and Chemical Model of Starless Cores of Magnetized Molecular Clouds. I. Formulation and Initial Results}",
      journal = {\apj},
         year = 2002,
        month = apr,
       volume = {569},
       number = {2},
        pages = {792-802},
          doi = {10.1086/339353},
archivePrefix = {arXiv},
       eprint = {astro-ph/0201019},
 primaryClass = {astro-ph},
       adsurl = {https://ui.adsabs.harvard.edu/abs/2002ApJ...569..792L}
}

@ARTICLE{1969MNRAS.144..425P,
       author = {{Penston}, M.~V.},
        title = "{Dynamics of self-gravitating gaseous spheres-III. Analytical results in the free-fall of isothermal cases}",
      journal = {\mnras},
         year = 1969,
        month = jan,
       volume = {144},
        pages = {425},
          doi = {10.1093/mnras/144.4.425},
       adsurl = {https://ui.adsabs.harvard.edu/abs/1969MNRAS.144..425P}
}

@ARTICLE{1969MNRAS.145..271L,
       author = {{Larson}, Richard B.},
        title = "{Numerical calculations of the dynamics of collapsing proto-star}",
      journal = {\mnras},
         year = 1969,
        month = jan,
       volume = {145},
        pages = {271},
          doi = {10.1093/mnras/145.3.271},
       adsurl = {https://ui.adsabs.harvard.edu/abs/1969MNRAS.145..271L}
}

@ARTICLE{2017OAst...26..285D,
       author = {{Dudorov}, Alexander E. and {Khaibrakhmanov}, Sergey A.},
        title = "{Hierarchical structure of the interstellar molecular clouds and star formation}",
      journal = {Open Astronomy},
         year = 2017,
        month = dec,
       volume = {26},
       number = {1},
        pages = {285-292},
          doi = {10.1515/astro-2017-0428},
       adsurl = {https://ui.adsabs.harvard.edu/abs/2017OAst...26..285D}
}

@ARTICLE{1976ApJ...207..484W,
       author = {{Woodward}, P.~R.},
        title = "{Shock-driven implosion of interstellar gas clouds and star formation.}",
      journal = {\apj},
         year = 1976,
        month = jul,
       volume = {207},
        pages = {484-501},
          doi = {10.1086/154516},
       adsurl = {https://ui.adsabs.harvard.edu/abs/1976ApJ...207..484W}
}

@ARTICLE{2004RvMP...76..125M,
       author = {{Mac Low}, Mordecai-Mark and {Klessen}, Ralf S.},
        title = "{Control of star formation by supersonic turbulence}",
      journal = {Reviews of Modern Physics},
         year = 2004,
        month = jan,
       volume = {76},
       number = {1},
        pages = {125-194},
          doi = {10.1103/RevModPhys.76.125},
archivePrefix = {arXiv},
       eprint = {astro-ph/0301093},
 primaryClass = {astro-ph},
       adsurl = {https://ui.adsabs.harvard.edu/abs/2004RvMP...76..125M}
}

\end{document}